# Shear thickening and migration in granular suspensions

Abdoulaye Fall[1,2], Anaël Lemaître[1], François Bertrand[1], Daniel Bonn[2,3], Guillaume Ovarlez[1]

[1] *Université Paris Est, Laboratoire Navier (CNRS UMR 8205), 2 allée Kepler, 77420 Champs sur Marne, France*
[2] *Van der Waals-Zeeman Institute, University of Amsterdam, Valckenierstraat 65, 1018XE Amsterdam, The Netherland*
[3] *Laboratoire de Physique Statistique, Ecole Normale Supérieure, 24 Rue Lhomond, 75005 Paris, France*

We study the emergence of shear thickening in dense suspensions of non-Brownian particles. We combine local velocity and concentration measurements using Magnetic Resonance Imaging with macroscopic rheometry experiments. In steady state, we observe that the material is heterogeneous, and we find that that the local rheology presents a continuous transition at low shear rate from a viscous to a shear thickening, Bagnoldian, behavior with shear stresses proportional to the shear rate squared, as predicted by a scaling analysis. We show that the heterogeneity results from an unexpectedly fast migration of grains, which we attribute to the emergence of the Bagnoldian rheology. The migration process is observed to be accompanied by macroscopic transient discontinuous shear thickening, which is consequently not an intrinsic property of granular suspensions.

The realm of complex fluids encompasses biological liquids such as blood, many liquid foodstuffs, building materials, glasses and plastics, crude oil, etc. Despite their importance, the most basic, quintessential question about the flow of these fluids has remained unanswered: it is generally impossible to predict their flow resistance, and it is even unclear why most fluids shear thin, whereas only some shear thicken. Understanding shear thickening, *i.e.* the increase of the apparent viscosity of materials with increasing flow rate, is thus an important issue in complex fluids with in addition a strong impact on energy consumption in industrial processes [1]. It is observed in dense colloidal suspensions [1,2], where it has been related to the formation of dense clusters of particles [2-4]. The viscosity rise with the shear rate is then usually reversible (it is a steady-state property), continuous, and is sharper at higher volume fractions [2-4]. For colloids, the competition between shear-induced cluster formation and Brownian motion that homogenizes the suspensions then naturally determine a critical shear rate for the onset of shear thickening.

As Brownian motion is absent in pastes made of large particles, the sharp shear thickening transition observed in for instance cornstarch suspensions [5] is highly surprising. In fact, the conditions of emergence of shear thickening in non-Brownian suspensions remain ill-characterized: in some systems, thickening was observed at low shear rates [1,5-7], while in others no shear thickening (only viscous behavior) is observed whatsoever, even close to jamming [8-10]. Up to now it is thus impossible to predict whether a given system will shear thicken or not.

In systems for which it is observed, a more pronounced shear thickening [6,7] is observed near jamming, similarly to colloidal suspensions, and is attributed to aggregation of hydroclusters into a percolating network [4,11]. However, one should question whether these observations of sharp and discontinuous shear thickening reflect an intrinsic (local, steady-state) property of materials. For example, an important effect of confinement on shear thickening was recently evidenced in both colloidal [12] and noncolloidal [5] suspensions. In channel flows [12], the macroscopic shear thickened state was shown to form a plug flow and was not observed in large channels, which shows that local observations are crucial to get a better insight into shear thickening.

In this Letter, we address these puzzles by studying the emergence of shear thickening in the simplest of systems: model density-matched suspensions of non-Brownian particles in water. We use a wide gap Couette geometry to avoid confinement effects, and we access the intrinsic material behavior by measuring the local flow properties and particle concentration using Magnetic Resonance Imaging (MRI). In steady state, we show that the material is heterogeneous, and that the *local* rheology presents a *continuous* transition, from a viscous to a shear thickening, Bagnoldian, behavior (shear stresses proportional to the shear rate squared) at any fixed volume fraction, as predicted by a scaling analysis. The heterogeneity is shown to result from an unexpectedly fast migration of grains during the transient, which is attributed to the emergence of the Bagnoldian rheology. The migration process is accompanied by *macroscopic* transient *discontinuous* shear thickening, which is thus not an intrinsic property of granular suspensions.

*Materials and methods* — We study dense suspensions of noncolloidal monodisperse spherical particles immersed in a Newtonian fluid. We use polystyrene beads (diameter 40 μm, polydispersity <5%, density 1.05 g.cm$^{-3}$) suspended in aqueous solutions of *NaI* to match the solvent and particle densities; the solution viscosity is 1 mPa.s. The density matching ensures that there are no gravity-induced contacts [13] and that the only source of normal stresses is shear [14,15]. We focus on results obtained at a 59% mean volume fraction as experiments at other high volume fractions show similar features. The material behavior is studied with a wide-gap Couette rheometer (inner radius: 4.1 cm; outer radius: 6 cm; height of sheared fluid: 11 cm) inserted in a MRI scanner, allowing us to access local velocity and particle volume fraction profiles in the flowing sample [9,16,17]. Sandpaper is glued to the walls and there is no significant slip on the velocity profiles. The inner cylinder velocity is controlled, and we record the torque exerted by the material on the inner cylinder with a Bohlin rheometer.

*Macroscopic vs. local behavior* — We first focus on the macroscopic behavior (Fig. 1a). The torque $T$ values measured during a slow ramp in rotational velocity $\Omega$ (logarithmic ramp, 45 s/decade of shear) are shown as black circles. At the beginning of the ramp, $T$ increases linearly with $\Omega$, as expected for a homogeneous, Newtonian suspension [9,10]. Around $\Omega_c \approx$ 2.5 rpm (corresponding to a low average shear rate of 0.6 s$^{-1}$), $T$ presents a sudden increase – by a factor of 20 – which is a usual signature of "discontinuous" shear thickening, following which $T$ continues to increase with $\Omega$, yet at a slower pace. At the end of the ramp, $\Omega$ is kept constant and $T$ reaches a stationary value. Subsequent slow changes (down or up) of the rotational velocity then drive the system reversibly along the curve $T(\Omega)$ in open symbols, *i.e.* the system has reached steady state. This curve presents very smooth, moderate, "continuous" shear thickening.

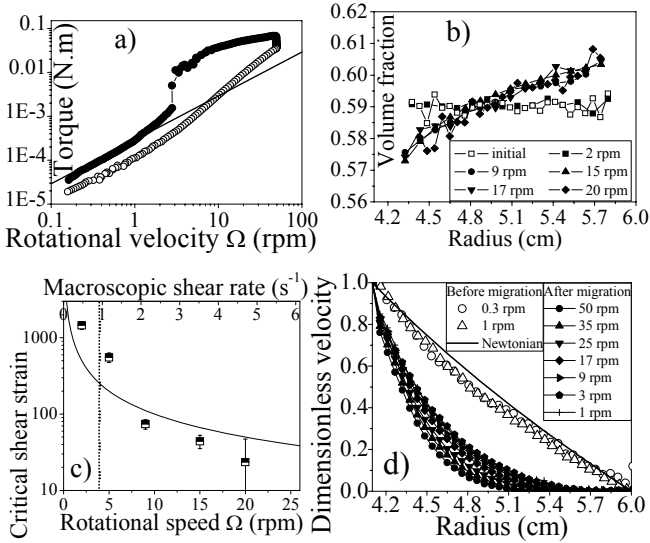

Fig. 1: a) Torque vs. rotational velocity $\Omega$ when shearing a 59% suspension: increasing velocity ramp (filled circles) and stationary state (empty circles); the line is a viscous law. b) Volume fraction vs. radius $R$, for various $\Omega$, during an increasing velocity ramp. c) Critical strain needed to complete migration in a 59% suspension vs. $\Omega$; the dotted line indicates the $\dot\gamma/\dot\gamma^2$ transition on the 59% suspension. The line is a $1/\dot\gamma$ scaling. d) Dimensionless velocity vs. $R$ (empty symbols: before migration, filled symbols: various $\Omega$ after migration, line: Newtonian velocity profile).

We now turn to local measurements. During the initial increasing velocity ramp, at a low rotational velocity, $\Omega$=2 rpm (<$\Omega_c$), the density $\phi(R)$ remains uniform (Fig. 1b) while the flow (Fig. 1d) is homogeneous– there is no jammed region [5]. The velocity profile $V(R)$ closely matches that of a Newtonian fluid (Fig. 1d) [9,18], consistent with the initial linear behavior of the torque. Shortly after shear thickening occurs (here at $\Omega$=9 rpm) we observe that the material has become strongly heterogeneous (Fig. 1b) while the velocity profiles present a jammed region near the outer cylinder. This change is irreversible: the density profiles subsequently remain similar, even when $\Omega$ is increased further and, later, decreased below $\Omega_c$.

Clearly, the discontinuous shear thickening observed during the initial up-ramp is a transient phenomenon associated with a large-scale reorganization of the material, which involves *shear-induced migration* from low to high shear zones. While migration is expected in dense suspensions [9,19-21], it is particularly striking here that the change in $\phi(R)$ occurs over a very short time interval, corresponding to a small total strain of order 100. Such a rapid migration is a puzzle as it is not predicted by classical theories [9].

*Constitutive behavior* — We now analyze the steady-state behavior. We first note that density and velocity profiles are in steady state whenever the torque is. Moreover, while $\phi(R)$ is $\Omega$-independent in steady state, the dimensionless velocity profiles $V(R,\Omega)/V(R_i,\Omega)$ measured at various $\Omega$ do not superpose, implying that the local behavior is not simply viscous [9,18]. Finally, the flow is always strictly localized: for all $\Omega$, there is a jammed region beyond a critical radius $R_m$=5.7 cm; this corresponds to density threshold $\phi(R) > \phi_m \approx$60.5% above which the material is jammed [9].

The material and flow being heterogeneous, macroscopic torque measurements $T(\Omega)$ are not sufficient to infer the intrinsic constitutive behavior, in particular the stress/strain-rate relationship in the shear thickening regime. This intrinsic behavior can however be obtained using our local measurements. The key point [9] is that the steady-state density profile $\phi(R)$ is independent of $\Omega$; a change of variables can then be performed between radius $R$ and $\phi(R)$. In addition, the stress profile is prescribed by momentum balance $\tau(R) = T/(2\pi H R^2)$ while the local shear rate can be extracted from the velocity profile $V(R)$ via: $\dot\gamma(R) = R\, \mathrm{d}(V/R)/\mathrm{d}R$. A local stress/strain-rate curve $\tau(\dot\gamma, \phi)$ – at fixed and well-defined density $\phi$ – is then obtained by collecting all measurements of local stress $\tau(R)$ and shear rate $\dot\gamma(R)$ for a fixed $R$ and varying $\Omega$.

The results of this local analysis (Fig. 2a) show that, for a fixed volume fraction $\phi$, a clear transition from a $\tau \propto \dot\gamma$ (Newtonian) to a $\tau \propto \dot\gamma^2$ (Bagnoldian) regime occurs at a critical shear rate $\dot\gamma_c(\phi)$ (Fig. 2b). Such a transition has been predicted to be a generic property of noncolloidal suspensions on the basis of theoretical dimensional arguments [22,23]. The $\dot{\gamma}^2$ scaling signals a regime where particle inertia dominates over viscous forces [8,23], leading to a behavior analogous to that of dry granular materials (it needs not be associated with collision-dominated flows as Bagnold suggested [24]). It is particularly striking that inertial scaling arises in our dense, highly damped, suspension, with particles of size of only ~10 μm. Moreover, the critical shear rate $\dot{\gamma}_c$ (i) is rather low (of order 1 s$^{-1}$), (ii) vanishes almost linearly as the volume fraction tends to $\phi_c \approx 60.5\%$, which (iii) is identical – within the experimental accuracy – to the threshold $\phi_m$ at which the material jams.

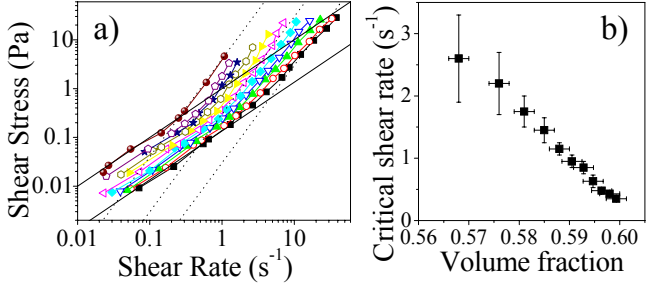

Fig. 2: a) Local shear stress vs. local shear rate measured for various local volume fractions when varying the inner cylinder rotational velocity from 0.1 to 50 rpm (from right to left: $\phi$=56.8%, 57.6%, 58.1%, 58.5%, 58.8%, 59%, 59.3%, 59.5%, 59.7%, 59.8%, 60%). The full lines are $\dot{\gamma}$ scaling; the dotted line are $\dot{\gamma}^2$ scaling. b) Critical shear rate for the $\dot{\gamma}/\dot{\gamma}^2$ transition vs. volume fraction.

*Viscous/inertial transition* — To understand what controls these scaling regimes, following [22,23], we write Newton's equations for a set of rigid particles. For the particles' centers of mass $r_i$ these read: $m\, d^2 r_i / dt^2 = \sum_j F_{ij} + F_i^{visc}$ where $F_{ij}$ denote rigid contact forces and $F_i^{visc}$ hydrodynamic forces which we suppose linear in terms of all the velocities entering into the problem. No other force is supposed to be involved. The key of the analysis is to remark that rigid forces $F_{ij}$ do not introduce, by definition, any force or length scale [22,23]. Two limiting cases can then be identified: "viscous" (V) when viscous forces are dominant over grain inertia $0 = \sum_j F_{ij} + F_i^{visc}$, and "inertial" (I) when grain inertia is dominant $m\, d^2 r_i / dt^2 = \sum_j F_{ij}$. Both expressions verify exact scale invariance by a change of time and force units [23], guaranteeing $F_{ij} \propto \dot{\gamma}$ in (V) and $F_{ij} \propto \dot{\gamma}^2$ in (I), with identical scaling with $\dot{\gamma}$ for all components of the stress tensor as rigorously shown in [23]. The full problem then reduces to (V) (resp. (I)) at low (resp. high) $\dot{\gamma}$, which explains the existence of a crossover between the two simple scaling regimes $\tau \propto \dot{\gamma}$ (viscous) and $\tau \propto \dot{\gamma}^2$ (inertial).

This formalism now helps us understand why the critical shear rate $\dot{\gamma}_c(\phi)$ can be so low and vanishes precisely at $\phi_m$. In the viscous (V) and inertial (I) regimes, the stresses are respectively of the form $\tau = \eta_o \dot{\gamma} \Sigma_V(\phi)$ and $\tau = \rho d^2 \dot{\gamma}^2 \Sigma_I(\phi)$. Numerical simulations [25] indicate that $\Sigma_V(\phi)$ and $\Sigma_I(\phi)$ should diverge at the same (jamming) packing fraction $\phi_m$ and read $\Sigma_V(\phi) \propto (\phi_m - \phi)^{-\alpha_V}$, $\Sigma_I(\phi) \propto (\phi_m - \phi)^{-\alpha_I}$, where $\rho$ and $d$ are the particle density and diameter, and $\eta_o$ is the interstitial fluid viscosity. The cross-over between the viscous and inertial regimes is found by equating the two expressions for the stress, finally leading to $\dot{\gamma}_c(\phi) \propto (\eta_0 / \rho d^2)(\phi_m - \phi)^{\alpha_I - \alpha_V}$. Together with the values $\alpha_I = 2$, $\alpha_V = 1$ proposed in the literature [26,27], this equation explains, as observed, that $\dot{\gamma}_c(\phi)$ vanishes (i) linearly with $\phi$, (ii) at the jamming packing fraction $\phi_m$. Moreover, the crossover stress verifies $\tau_c(\phi) \propto (\eta_0^2 / \rho d^2)(\phi_m - \phi)^{\alpha_I - 2\alpha_V}$, which, together with the same values of $\alpha_I, \alpha_V$ as above, suggests that $\tau_c$ should indeed be independent of volume fraction. Although our stress measurements are not sufficiently accurate to assert that $\tau_c(\phi) \sim$ cst, we then note in Fig. 2a that, indeed, in the experiments $\tau_c$ does not vary much. We finally conclude that it is the difference in singular behavior of the inertial and viscous stresses $\rho d^2 \dot{\gamma}^2 \Sigma_I(\phi)$ and $\eta_0 \dot{\gamma} \Sigma_V(\phi)$ at the approach of jamming (*i.e.* when $\phi \to \phi_m$) which leads to the linear vanishing of $\dot{\gamma}_c(\phi)$, and hence permits this transition to take place at low strain rates.

*Accelerated shear-induced migration* — We now show that this transition explains the sudden migration associated with the macroscopically observed transient discontinuous shear thickening (Fig. 1a). In viscous suspensions, shear-induced migration is usually thought to be negligible when small particles are involved. Indeed, the typical strainscale for migration is very large: it is expected to be rate independent and to scale as $\propto (R_o - R_i)^2 / a^2$ [9,19-21], leading to an expected strain of order 50000 [9], more than 500 times higher than what we observe here at the onset of shear thickening. Our observations may be understood as a strong enhancement of migration kinetics in the inertial regime.

Within the framework of two-phase models, migration is driven by gradients of internal *normal stresses* within the particle network $\Sigma_{ii}^p$ (not the total stress), and requires the fluid to filter through the granular phase to compensate for the local changes of packing fraction [20,21]. This filtration process exerts an average hydrodynamic drag $\propto U$ on the particle network, with $U$ the average filtration velocity. The balance between these two effects controls the migration/filtration rate, leading to $U \propto \nabla \Sigma_{ii}^p$. When injected in a mass conservation equation $\partial \rho / \partial t = -\nabla(\rho U)$, this leads to a diffusion equation for the particle density $\rho$ [20,21]. The local particulate stress $\Sigma_{ii}^p$ entering this analysis is expected to display local viscous or inertial scaling over very short strainscales [22,25,26] compared to those of the migration process. If $\Sigma_{ii}^p \propto \dot{\gamma}$ in the whole system, the timescale of migration scales as $1/\dot{\gamma}$ and migration is controlled by a (large) rate-independent strainscale, which is the classical result [19-21]. Strikingly, the same analysis performed in the inertial regime now yields an unexpected $1/\dot{\gamma}$ strainscale for migration: this explains why migration is much accelerated, and manifests itself abruptly when entering the inertial regime.

To confirm this analysis, we have studied the migration kinetics at constant $\Omega$'s, starting each time from a homogeneous state. We define the critical strain $\Gamma_{\text{migr}}(\Omega)$ as the macroscopic strain above which the instantaneous volume fraction profile matches the steady state one within experimental uncertainty (0.2 %). Fig. 1c shows $\Gamma_{\text{migr}}(\Omega)$ vs. $\Omega$: it decreases strongly with $\Omega$. Although the exact kinetics results from a complex history (as both $\dot{\gamma}$ and $\phi$ change locally in time), the asymptotic $1/\dot{\gamma}$ decay predicted by the above scaling analysis in the Bagnoldian regime is roughly consistent with our observations (see line in Fig. 1c). Migration theories based on normal stresses [20,21] are thus shown here to be more generally applicable than diffusive theories [19].

*Macroscopic shear thickening* — To summarize, we propose the following scenario: (i) the intrinsic behavior of dense noncolloidal suspension presents a continuous transition at low strain rates from a viscous to a shear thickening, Bagnoldian, rheology characterized by shear stresses $\propto \dot{\gamma}^2$; (ii) in the Bagnoldian regime, a very fast particle migration then occurs towards low shear zones; (iii) the interplay between flow and migration shows up as a sharp shear thickening of the transient macroscopic stress.

Note that we checked the robustness of the above scenario by verifying that it is not specific to wide gap Couette geometry. We have studied the behavior of the same material in a parallel plate geometry and have observed similar phenomenology, *i.e.* transient discontinuous shear thickening associated with fast migration due to the $\propto \dot{\gamma}^2$ scaling of stresses.

We here contrast our findings with those of Brown and Jaeger [7] in a similar system (spheres in a Newtonian fluid). On the basis of steady-state macroscopic measurements, they find a transition between a roughly viscous and a shear thickening regime $\tau \propto \dot{\gamma}^{1/n}$ with $n$ continuously decreasing for 0.5 to 0 when $\phi \to \phi_m$. Here, we find instead that the local (intrinsic) rheology shows only viscous or Bagnoldian scalings (*i.e.* $n$=0.5) even when $\phi \to \phi_m$. We insist that in steady state such systems are heterogeneous, and that consequently the macroscopic stress/strain-rate relationship cannot be directly related to the local constitutive behavior, in particular in the shear thickening regime.

Let us finally note that our mechanism may also be at work in Brownian suspensions, in competition with or as an alternative to hydrodynamic clustering [2]. It is compatible with the reversibility of the shear thickening transition usually observed in Brownian suspensions: migration is indeed expected to be reversible due to the osmotic pressure. It is therefore particularly striking that our mechanism leads to a constant critical shear stress, exactly what is observed for colloids [2].

We thank Daniel Lhuillier for enlightening discussions on shear-induced migration.